# Energetic Electrons Associated with Magnetic Reconnection in the Magnetic Cloud Boundary Layer

Y. Wang,[1,2] F. S. Wei,[1] X. S. Feng[1], S. H. Zhang[1,2], P. B. Zuo[1] and T. R. Sun[2]

1 SIGMA Weather Group, State Key Laboratory for Space Weather, Center for Space Science and Applied Research, Chinese Academy of Science, Beijing 100190, China

2 Graduate University of the Chinese Academy of Science, Beijing 100049, China

**Abstract** Here is reported in situ observation of energetic electrons (~100-500 keV) associated with magnetic reconnection in the solar wind by the ACE and Wind spacecraft. The properties of this magnetic cloud driving reconnection and the associated energetic electron acceleration problem are discussed. Further analyses indicate that the electric field acceleration and Fermi type mechanism are two fundamental elements in the electron acceleration processes and the trapping effect of the specific magnetic field configuration maintains the acceleration status that increases the totally gained energy.

Magnetic reconnection is a process that quickly converts magnetic energy into thermal and kinetic energy [1]. It acts as an efficient accelerator to produce energetic electrons that are always found in the solar corona and terrestrial magnetosphere [2-5]. Observations have indicated that up to 50% of the energy can be released to accelerate 20-100 keV electrons during the reconnection in the solar flares [3]. Direct measurement of electrons accelerated up to 300 keV in the magnetic reconnection diffusion region of the earth's magnetotail has also been reported [4]. Although simulations have predicted the possible acceleration of ~100 keV particles in the sector boundaries [6], no such energetic electrons have so far been found in the solar wind reconnection [7-11]. It is very important to know why they are missed in these reconnections and how they are produced.

Early test particle analyses based on magnetohydrodynamic (MHD) simulations indicate that the presence of turbulent fluctuations can help accelerate particles to high-energy by trapping them in the strong electric field region [6, 12-13]. This turbulent mechanism includes both coherent and stochastic components of acceleration. Recent particle-in-cell (PIC) simulations also suggest that both the direct electric field acceleration near the X-line and the Fermi-like mechanism in the contracting magnetic islands can generate electrons up to a few hundred keV under certain circumstances [14-17]. The key of the former model is the intensity of electric field as well as the effective acceleration time, and the soul of the later one requires volume filled conctracting magnetic islands. However, up to now, no special acceleration model has ever been proposed for the solar wind case and the physical mechanism responsible for producing energetic electrons is still a controversial topic.

In this letter, we present the first observation of energetic electrons (~100-500 keV) associated with magnetic reconnection in a magnetic cloud boundary layer (MCBL) [18-21]. Simulations based on this event suggest that the generation of energetic electrons in the solar wind is a combined process dominated by both the reconnection electric field and the Fermi type reflection; the large-scale and quasi-steady properties of the magnetic cloud (MC) driving reconnection are the foundation and precondition that provide electrons with sufficient acceleration time and space; the trapping effect of the specific magnetic field configuration maintains the acceleration status that boosts the finally reached energy.

On 3 October 2000, the Wind spacecraft measured a solar wind reconnection event at x, y, z=[32.6, -251.7, -3.8]Re (earth radii) in geocentric solar ecliptic (GSE) coordinates [Fig.1]. The region marked between Mf and Gf, where the MC interacts with ambient solar wind, is the MCBL [18-21]. Inside this area, a reconnection exhaust (labeled by A1 and A2 in Fig.1 and Fig.2) was identified by constructing the current sheet (LMN) coordinate system and verifying the Walen relation [22-25]. The magnetic field across the exhaust rotated ~143° with a nearly constant zero normal field ($B_N$) and an entirely reversed antiparallel field ($B_L$). The antiparallel velocity ($V_L$) in the exhaust was just consistent with the Alfvénic reconnection jet, and the small normal velocity shear ($\triangle V_N$=12 km/s) across bifurcated current sheet could be considered as a 6 km/s ($V_{in}=\triangle V_N/2$) reconnection inflow. Correspondingly, the reconnection electric field and the dimensionless reconnection rate were calculated to be 0.084 mV/m and 0.053 respectively [1]. Thanks to the ACE observation of the same event 2 hours later, we reconstructed and sketched this event in Fig.2(a) by assuming a symmetric geometric configuration. Due to the 14 minutes' passage with a 300 km/s velocity ($V_N$), the exhaust width obtained from ACE was ~$2.5\times10^5$ km (or 40 Re) and the distance to the X-line was ~$2.4\times10^6$ km (or 374 Re). Nevertheless, it seems impossible to obtain the real length of the X-line since its direction (*M*) is nearly perpendicular to the ecliptic plane (*M*=[-0.02$n_x$, -0.06$n_y$, 0.99$n_z$], calculated by minimum variance analysis of the magnetic field [24]) while both spacecrafts are almost in this plane. However, the observed similar characteristics could still indicate that this Petschek-type reconnection with quasi-steady and large-scale properties had a long X-line extending several hundred of earth radii as those previously reported observations [9-10] and this point will be also supported by the following simulations.

In this event, we fortunately observe energetic electrons associated with magnetic reconnection that have not been so far reported in the solar wind. As shown in Fig.3, unlike some of the previous energetic electron observations, the present spectrum displays strong pitch angle anisotropy and energy dependence. In the high-energy range (100-500 keV) where our greatest concern exists, the electrons have a burst increase in the antiparallel field direction near 16:55UT but seem virtually unchanged in the parallel field direction. The least square fit of these electrons, measured by Wind (SST-Foil), also shows power law distribution ($f(v) \propto E^{-k}$) with a smaller value of the power *k* inside the exhaust (at 16:55UT, *k*=3.7) and larger ones outside (at 16:42UT, 17:02UT, *k*=3.9, 3.8 respectively). Although the spacecraft does not cross the reconnection diffusion region, where the energetic electrons are thought to be produced [14-17], we can still rule out the possibility that such irregular flux enhancement is merely caused by the variation of background electron temperature or density (see the isotropic low-energy electrons around 40eV for example). Instead, both the reconnection topological structure and the harder spectrum in the exhaust suggest that they are most probably accelerated by magnetic reconnection and then expelled through the long reconnection separatrices [11, 14-17]. If so, since the local magnetic field at 16:55UT just points to the reconnection region and the expelled energetic electrons would mainly concentrate near specified two of the four reconnection separatrices by the modulation of the guide field [26], the Wind spacecraft would naturally measure enhanced field-aligned energetic electrons in the antiparallel direction [Fig.3(c), (e)].

As indicated in the beginning, how theses energetic electrons are produced in the magnetic reconnection is a long standing problem. Different local conditions will lead to different reconnection evolutions and affect the acceleration picture. Therefore, our first effort is to use the 2.5-dimensional MHD parallel adaptive mesh refinement (AMR) simulation to reveal the evolution of this MC driving reconnection [19, 27] under real solar wind conditions and then discuss the possible effects on the electron energization. The simulation [Fig.4(a)] suggests that this event is essentially the same as those reported solar wind Petschek-type reconnections [9-10] that own the quasi-steady and large-scale properties. The driving of the MC, a distinct feature of this event, results in an increased guide field [Fig.2(b), (e)] and reconnection electric field. Turning to the acceleration problem itself, from the global perspective, in this type of reconnection where a long-extend X-line exists, the enhanced reconnection electric field would bring about a large amount of reconnection potential energy. Then, focusing on the reconnection region, a compressed long thin current sheet with a prominent guide field, externally driven flows and high Lundquist number (the ratio of the time scales of resistive diffusion to typical Alfvén waves) conditions prefers to generate magnetic island chain rather than form a single X-line [2, 28-29]. In such cases, the appreciable energy gained from the contracting magnetic islands as the Fermi type mechanism should not be neglected either. Therefore, the above discussions imply that the electron energization in this event ought to be a combined process dominated by both the reconnection electric field and the Fermi type mechanism.

To see a detailed electron acceleration picture, we carry out another MHD computation using initial conditions similar to the geospace environmental modeling (GEM) reconnection challenge with a guide field [30], and based on this computation the test particle approach is also applied (see Fig.4 caption for details). From our results [Fig.4(b)], some electrons are able to be accelerated to high-energy if they are trapped by the specific magnetic field configuration. Particularly, Fig.4(b), taken at time t=151$t_A$ ($t_A$ is the Alfvén transit time, 1$t_A$=0.53 s) when obvious magnetic island chain (dashed lines) forms, shows a complicated trajectory of a trapped electron (solid line). This test electron starts at L, M, N=[2.0, 0.0, 0.7]$d_i$ ($d_i$ is the ion inertial length, 1$d_i$=73 km) with an initial energy of 14.4 keV and pitch angle 10°. As shown in Fig.4(c), the test electron does not get energy all the time. Specifically, significant energizations mainly occur at three typical acceleration stages (marked by colored rectangles respectively). In the beginning, the electron bounces in the middle island [Fig.4(b)] and gets energy at the contracting island ends from both the Fermi type reflections and the electric field [6, 12-17]. Later, the electric field acceleration continues as the electron takes Speiser-type motion across the current sheet to the right island and mirrors back in the magnetic field pileup region [15-16]. During the trapping status, the electron switches among the three typical motions and boosts its energy by both the Fermi type acceleration and the electric field until it totally drifts out. It is noteworthy that even if the island contraction is throttled by the back pressure of the heated electrons and the large guide field [14], the electron can still get reconnection potential energy by drifting in the X-line direction. In such a framework, it can be finally heated up to 500 keV in 30 seconds with a requirement of drifting 600 Re [Fig. 4(d)]. These results are consistent well with this event that the gained energy matches the observations and the acceleration time as well as the required long X-line is also acceptable in this type of reconnection [9-10]. Furthermore, although not observed here, the accelerated energetic electrons that peak both inside the magnetic islands and the current sheet in the simulation are also accordant with the related geomagnetic observation [2].

The traditional energetic electrons acceleration models provide reasonable explantions under their favored local conditions [6, 12-17]. However, the single X-line configuration assumed by the electric field acceleration model might not exist in some cases, and the islands volume filled structure proposed by the Fermi type mechanism might also have difficulty in explaining why the expelled energetic electrons are always concentrated near the reconnection separatrices [1-2, 5, 28-29]. More importantly, it might be too idealized for the referred models to treat acceleration of energetic electrons as a single mechanism dominated process [14-17], especially in the solar wind. Referring to this event again, the weak electric field itself is not able to produce electrons up to 500 keV alone since the maximum reconnection potential energy is only 317 keV even if the X-line could extend as long as in the simulation. On the other hand, the released energy from the contracting magnetic island prevented by both the large guide field and the electron back pressure is unlikely to independently generate such high-energy electrons either. Based on the observations, our simulations clearly conclude that the generation of energetic electrons in the solar wind is a combined process controlled by both the reconnection electric field and the Fermi type mechanism, and the trapping effect of the multi-islands configuration maintains the acceleration status that boosts the finally reached energy.

To sum up, energetic electrons are not always observed with the occurrence of magnetic reconnection and the referred acceleration models [6, 12-17] might be unsuitable to explain the generation of these electrons under real solar wind conditions. However, our observations extend the previous work into the interplanetary space and magnetic reconnection is demonstrated to be able to generate energetic electrons once again. Moreover, the proposed combined model might bring a new sight on the electron acceleration problem and it could also be applied in other similar cases such as the far magnetotail. Certainly, although the simulation results correspond well with the related observations, more self-consistent computations together with detailed observations inside the reconnection diffusion region should be provided. Comprehensive researches on the behaviors of magnetic reconnection under various conditions may have great implications to solve the general energetic electron acceleration problem and help us understand the essence of the energy conversion.

## Acknowledgements

We thank W. Matthaeus for the valuable suggestions. The data used in this paper are from the Wind and ACE spacecraft. This work is jointly supported by the National Natural Science Foundation of China (40890162, 40904049, 40921063, and 41031066), 973 program under grant 2006CB806304, and the Specialized Research Fund for State Key Laboratories.

**Figure Caption**

FIG.1 (color) (a)-(f) Wind measurements of density, magnetic field, velocity ($V_x$ has been shifted by 500 km/s) and temperature between 16:00-18:00 UT on 3 October 2000 with a cadence of 3 seconds. The MCBL is shown by the dot lines; the dashed lines marked by A1 and A2 inside the MCBL represent the reconnection exhaust boundaries at 16:46UT and 16:58UT respectively.

FIG.2 (color) (a) Geometric configuration of the reconnection event; the reconnection jets and X-lines are shown in red; the following MC is sketched in green. (b)-(d) ACE measurements of magnetic field and velocity in the LMN coordinate system ($V_L$ has been shifted by -500 km/s). The dashed lines stand for the exhaust boundaries. (e)-(g) Wind measurements of magnetic field and velocity in the LMN coordinate system ($V_L$ has been shifted by -500 km/s). The black solid lines show the predicted velocity. In the predication, we calculate the theoretical velocity by the Walen relation: $\mathbf{v}_{pre} = \mathbf{v}_{ref} \pm \rho_{ref}^{1/2}(1-\alpha_{ref})^{1/2}(\mathbf{B}/\rho - \mathbf{B}_{ref}/\rho_{ref})/\mu_0^{1/2}$ [P. D. Hudson, Planet Space Sci 18, 1611 (1970); G. Paschmann et al., J Geophys Res 91, 1099 (1986)]. Here V, B, ρ represent normal meanings, while $\alpha \equiv (p_{//} - p_{\perp})\mu_0/B^2$, the pressure anisotropy factor, is assumed to be zero. The subscript 'ref' denotes the reference time at the leading (trailing) edge of the exhaust and 'pre' denotes the calculated velocity across the region. The positive (negative) sign is chosen for the trailing (leading) edge of the exhaust. The leading and trailing edge predictions merge at 16:53UT.

FIG.3 (color) Wind measurements of 10 eV-500 keV electron normalized fluxes in parallel (a), perpendicular (b) and antiparallel (c) directions respectively. (d) 5-5000 keV proton normalized flux (omni). All the fluxes are divided by the background flux $F_0$ at every energy band for normalizing. The background flux $F_0$ is chosen as the mean flux of a nearby quiet period (between 16:00-16:30UT) for each energy band. The missing (invalid) data is shown by blank. (e) Normalized fluxes at 16:55UT.

FIG.4 (color) (a) MHD parallel AMR simulation of the MC driving reconnection in the solar wind (left) and the zoomed result (right). The magnetic field lines are sketched by the solid curves with arrows. Detailed simulation methods are given in F. S. Wei et al., Space Sci Rev 107, 107 (2003) and references therein. The radius of the MC projected in the LN plane is 300 Re and the whole computation region is L×N=2000 Re ×1200 Re. Other input parameters are chosen from Fig.1 and Fig.2. (b) MHD parallel AMR simulation of magnetic reconnection in the reconnection region. The magnetic field is sketched by the dotted lines. The initial simulation conditions similar to the GEM reconnection challenge with a guide field [J. Birn et al., J Geophys Res 106, 3715 (2001)]. The aspect ratio of the Harris sheet is set to 100 and the system size is L×N=100 $d_i$×10 $d_i$; the initial configurations include 6 km/s inward driven flows at both sides and a uniform 5 nT guide field. Other parameters are directly obtained from Fig.1 and Fig.2. (c) The electron energy $E_k$ as a function of its position in L direction. Colored rectangles mark the three typical motions. (d) The electron energy $E_k$ (blue) and its *M* position (red) as a function of time during the whole acceleration process.

Figure 1(color)

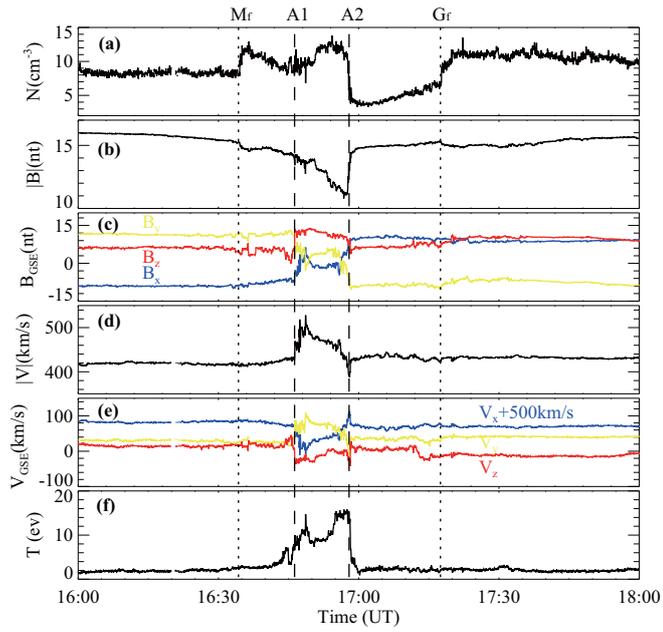

Figure 3(color)

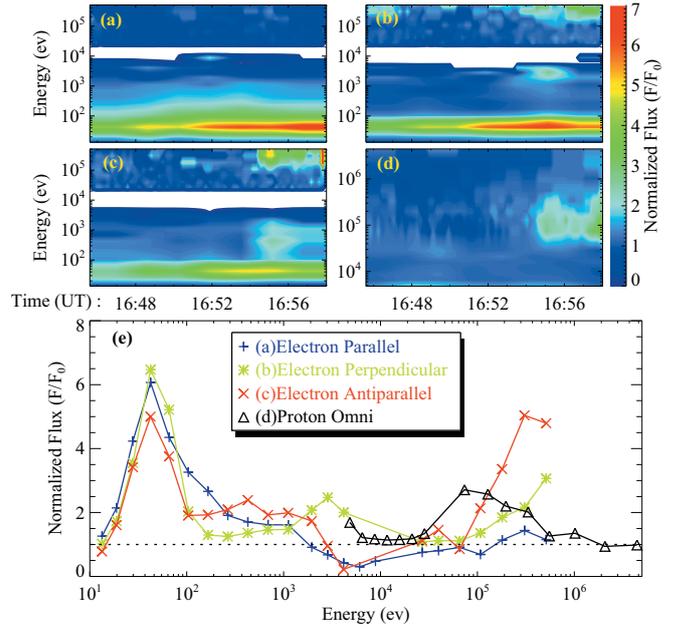

Figure 2(color)

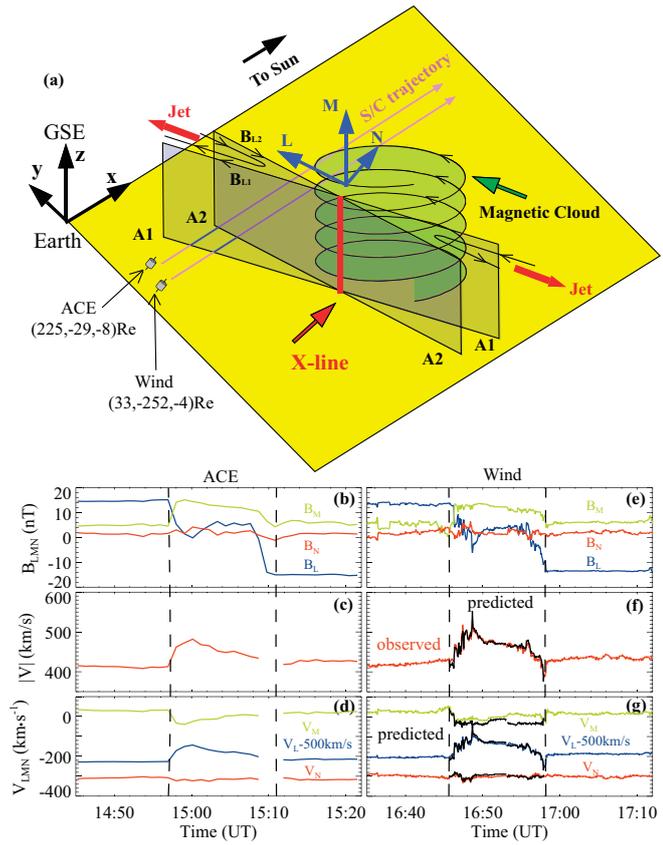

Figure 4(color)

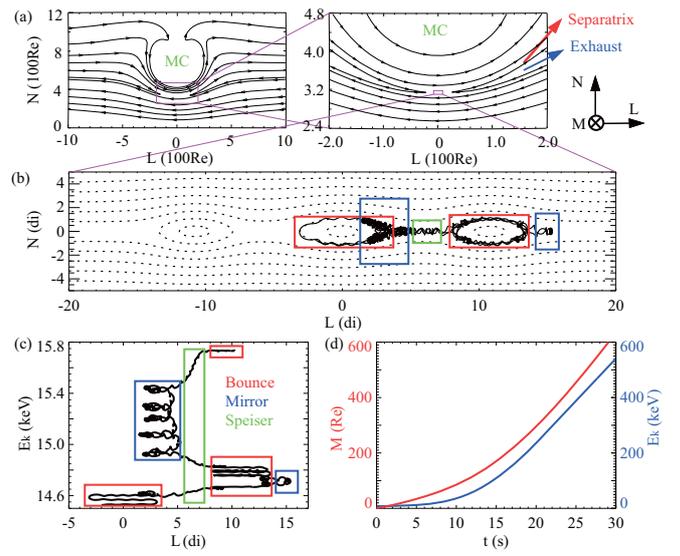